# Effect of non-Fourier heat transport on temperature distribution in High Bandwidth Memory

*Zhihao Zhou, Yu He, Shixian Liu, Lina Yang, Nuo Yang*

*Abstract*—**High Bandwidth Memory (HBM), as a key development trend in future memory chip technology, significantly enhances computer performance. At the same time, the thermal challenges arising from its stacked architecture have drawn considerable attention. Most existing studies on HBM thermal management are based on Fourier's law, neglecting the non-Fourier effects introduced by the micro/nanoscale structures within HBM. In this study, the Monte Carlo method (MC) is employed to solve the phonon Boltzmann transport equation (BTE) and investigate the impact of non-Fourier heat transport on the thermal behavior of HBM structures. The results reveal that non-Fourier heat transport leads to a junction temperature that is 59.8 °C higher than that predicted by Fourier's law. Furthermore, it is found that the phonon transmittance at the chip interlayers has a severe impact on heat dissipation, with the temperature variation reaching up to 56.6°C. These findings provide more accurate thermal insights, which are critical for the optimized design of HBM systems.**

Zhihao Zhou is with the School of Energy and Power Engineering, Huazhong University of Science and Technology, Wuhan 430074, China, and also with the School of Science, National University of Defense Technology, Changsha 410073, China (e-mail: zhihao_zhou@hust.edu.cn).

Yu He and Nuo Yang are with the School of Science, National University of Defense Technology, Changsha 410073, China (e-mail: heyu25@nudt.edu.cn; nuo@nudt.edu.cn).

Shixian Liu is with the Department of Thermophysics, Bauman Moscow State Technical University, Moscow 105005, Russia (e-mail:sxliu98@gmail.com ).

Lina Yang is with the School of Aerospace Engineering, Beijing Institute of Technology, Beijing 100081, China (e-mail: yangln@bit.edu.cn).

(Corresponding author: Lina Yang and Nuo Yang)

## I. INTRODUCTION

IN recent years, the rapid progress in artificial intelligence and big data technologies has significantly heightened the demand for high-performance memory systems [1], [2], [3]. Consequently, Joint Electron Device Engineering Council (JEDEC), the leading standardization organization for semiconductor memories, has introduced a cutting-edge solution known as high-bandwidth memory (HBM) [4]. HBM uses a 3D stacked architecture to achieve ten times higher bandwidth than traditional Double Data Rate 5 (DDR5), but it also brings more serious thermal problems [5], [6], [7]. HBM's smaller surface area, higher power density, and greater interlayer thermal resistance all deteriorate the heat dissipation of the chip [8], [9], [10], [11]. The resulting high temperature will directly affect the signal transmission and electrical stability of HBM [12], [13], [14]. Therefore, studying the internal temperature distribution of HBM is crucial for its further advancement and utilization.

The practical heat management issue in HBM has garnered significant attention. Kim et al. [15] found that the thermal resistance generated by stacking chips can lead to a temperature rise of up to 55 ºC. Zhang et al.'s study [16] indicated that the reduction of chip thickness to 1 $\mu m$ would further exacerbate thermal diffusion. To address the high temperatures in HBM, researchers have proposed various heat dissipation schemes, such as microfluidic cooling [17], embedded cooling [18–20], [19], [20], and jet impingement cooling [21]. These methods have been shown to reduce the chip temperature by several tens of degrees Celsius. Many efforts have been made to conduct thermal analysis on HBM using Fourier's law. However, as the size of certain structures in HBM decreases to the micro- and nanoscale, Fourier's law becomes inapplicable [22], [23], [24], [25] and Fourier's law-based thermal simulations may result in inaccurate thermal design rules [26], [27].

In the field of chips and transistors, due to their size, Boltzmann transport equation (BTE) is a popular model for efficient thermal analysis. Many scientific research institutions, such as Intel [28] and IMEC [29], have conducted extensive researches on the ballistic effect in transistors by iteratively solving phonon BTE. Over the past few years, Hua Bao's group [30] has developed GiftBTE and has made excellent progress in the self-heating topic of transistors [31], [32], [33]. They found that considering non-Fourier thermal transport in fin transistors would result in a temperature rise 40.5 ºC higher than that of Fourier's law [32]. However, compared to deterministic methods, the Monte Carlo (MC) method offers significant advantages in handling the complex structure of real chips, providing more flexibility and accuracy in simulations [34]. Cao et al. have conducted many in-depth studies on the hot spots of chips using the MC method [35], [36], [37], [38]. Hao et al., through electron-phonon MC simulation, found that Fourier's law significantly underestimates the temperature of hotspots in transistors [39], which approached 40 ºC.

In this work, the temperature distribution of HBM is numerically studied using phonon Monte Carlo simulations. Firstly, the size dependence of the temperature distribution is studied, and the impact of the non-Fourier effect on the temperature is evaluated by comparing the results of the Monte Carlo simulations based on both the grey and non-grey models with those predicted by Fourier's law. Subsequently, to investigate the influence of the joints on the heat dissipation of HBM, interlayer thermal resistance is introduced. The effect on temperature is analyzed by varying the phonon transmittance and specularity at the joints. This work provides valuable insights for thermal simulations and the thermal design of HBM and similar devices.

## II. METHOD AND VALIDATION

### A. Boltzmann transport equation

The Boltzmann transport equation is a commonly used model for heat conduction at the microscale and nanoscale [40], [41], [42], [43]. Compared with other small scale heat conduction modeling methods such as molecular dynamics simulations and ab initio calculations, BTE is suitable for a wider range of length scales [23], [34]. Although the phonon coherence is ignored, BTE can use limited computational resources to get its heat distribution accurate enough for the size of the chip [32], [37], [44]. In the BTE, phonon bundles are treated as particles, neglecting their wave properties, and it is used to describe the evolution of the single-particle probability distribution function $f(\mathbf{x},\mathbf{k},p,t)$ over time. BTE in the frequency-dependent relaxation-time approximation is generally expressed as [41]

$$\frac{\partial f}{\partial t} + \mathbf{V_g}(\omega,p)\nabla f = -\frac{f - f^{loc}}{\tau(\omega,p,T)} \quad (1)$$

where $\omega = \omega(k,p)$ is the phonon angular frequency, p is the phonon polarization, T is the temperature, $\mathbf{V}_g = \nabla_k \omega$ is the phonon group velocity and $\tau$ is the phonon relaxation time. Also, $f^{loc}$ is an equilibrium (Bose-Einstein) distribution parametrized by the local pseudotemperature. The equilibrium solution of Eq. (1) at temperature T is given by the Bose-Einstein distribution

$$f^{eq}(\omega,T) = \frac{1}{\exp(\frac{\hbar\omega}{k_b T}) - 1} \quad (2)$$

where $k_b$ is Boltzmann's constant.

When the conventional MC method is used to solve Eq. (1), the scattering process does not strictly satisfy energy conservation [43], [46]. To solve this problem, multiply Eq. (1) by $\hbar\omega$ and define $e = \hbar\omega f$ and $e^{loc} = \hbar\omega f^{loc}$ to obtain the energy-based BTE

$$\frac{\partial e}{\partial t} + \mathbf{V_g}(\omega,p)\nabla e = -\frac{e - e^{loc}}{\tau(\omega,p,T)} \quad (3)$$

In this formulation, each computational particle represents a fixed amount of energy $e = \hbar\omega f$ and a strict energy conservation is achieved by conserving the number of particles. In addition, the conventional MC simulation is a low signal-to-noise (S/N) ratio when departure from the equilibrium distribution is small [47]. The variance reduction method can effectively reduce the influence of statistical noise on the result. By defining an equilibrium energy distribution $e^{eq}_{T_{eq}} = \hbar\omega f^{eq}(\omega,T_{eq})$ as a control variate, the deviation from equilibrium $e^d = e - e^{eq}_{T_{eq}}$ can be solved. Eq. (3) can be recast as [37], [44]

$$\frac{\partial e^d}{\partial t} + \mathbf{V_g}(\omega,p)\nabla e^d = -\frac{e^d - (e^{loc} - e^{eq}_{T_{eq}})}{\tau(\omega,p,T)} \quad (4)$$

The choice of $T_{eq}$ should satisfy $(T - T_{eq}) \ll T_{eq}$, control variate $e^{eq}_{T_{eq}}$ is close to the actual distribution $e$, thus providing an ideal condition for variance reduction.

### B. Monte Carlo solution of BTE

Phonon MC method is a stochastic method to solve phonon BTE, which can effectively solve the heat transport problem of complex geometry and arbitrary heat source distribution [49], [50], [51], [52]. For deviational energy-based BTE, MC method generates samples from the initial deviational energy distribution $e^d(\mathbf{x} = 0, t = 0)$. The samples arrive at a new location through drifts and scatterings, and a new distribution $e^d(\mathbf{x},t)$ is obtained by statistics on sample in each $(\mathbf{x},t)$. Phonon-phonon and phonon-boundary scattering are considered in the model. Here, $e^d$ is sampled by N computational particles using

$$e^d(\mathbf{x},t,\omega,p,\theta,\phi)\frac{D(\omega,p)}{4\pi} = \\ \varepsilon^d_{eff} \sum_i s_i \delta^3(\mathbf{x}-\mathbf{x_i})\delta(\omega-\omega_i)\delta(\theta-\theta_i)\delta(\phi-\phi_i)\delta_{p,p_i} \quad (5)$$

where $s_i$ is the sign of a computational particle given by the sign of $e^d = e - e^{eq}_{T_{eq}}$. If the deviational energy is positive, a positive particle is produced, and vice versa a negative particle is produced. $D(\omega,p)$ is the density of phonon states, given by

$$D(\omega,p) = \frac{k(\omega,p)^2}{2\pi^2 V_g(\omega,p)}.$$

The total deviational energy of the system is calculated by combining the contributions of all the sources. Abhishek Pathak et al. [53] derived the contribution of the initial conditions and isothermal boundary to the deviational energy. According to their method, we add a volumetric heat source [32], [38]

$$Q_{init} = \delta(t)(T_{init} - T_{eq})\frac{de^{eq}_{T_{eq}}}{dT}$$

$$Q_{isothermal} = \delta(\mathbf{x})H(\mathbf{V_g}\cdot\hat{\mathbf{n}})(\mathbf{V_g}\cdot\hat{\mathbf{n}})(T_b - T_{eq})\frac{de^{eq}_{T_{eq}}}{dT} \quad (6)$$

$$Q_{heatflux} = q_{\omega,p}$$

Here, $T_{init}$ is the initial temperature, $H$ is the Heaviside function specifying the direction of particle emission, $\hat{n}$ is the inward normal to the isothermal boundary, and $T_b$ is the boundary temperature. The deviational energy associated with the *ith* source is given by

$$E^d_i = \sum_p \int_t \int_\omega \int_\phi \int_\theta \int_V \frac{D}{4\pi}|Q_i|dV\sin(\theta)d\theta d\phi d\omega dt \quad (7)$$

where, $Q_i$ is the phase-space energy density associated with the *ith* source, and $dV$ is the differential volume element in the position space. The total deviational energy $E^d_{tot}$ is obtained by summing the deviational energy of all sources in the system. The deviational energy of phonon bundle is calculated as $\varepsilon^d_{eff} = E^d_{tot}/N$, where N is the number of phonon bundles. In this study, N is 50 million to ensure the accuracy of the simulation.

For the entire work, first, first-principles calculations based on density functional theory (DFT) and density functional perturbation theory (DFPT) were employed to obtain the interatomic force information of the relaxed crystal structure. Harmonic (2nd) force constants were extracted using DFPT by computing the linear response of the system to atomic displacements. Anharmonic (3rd) force constants, the finite displacement method was used, where a series of supercell configurations with systematically displaced atoms were generated, and the resulting forces were calculated using DFT. Secondly, these force constants are used as the input of ShengBTE to obtain phonon properties, including group velocity, relaxation time, and heat capacity. Finally, taking phonon properties as input, the phonon transport in HMB is simulated using the MC method. The flowchart is shown in Figure 1.

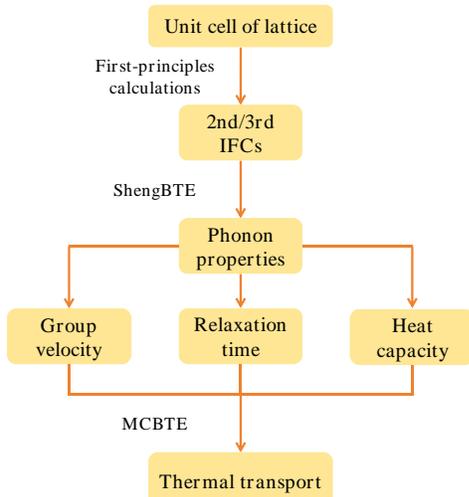

Fig. 1. The calculation flowchart of the Monte Carlo method based on first principles calculation.

## C. Validation of the algorithm of the heat flux source

To verify the correctness of the algorithm of heat flux source, a typical internal heat generation problem is calculated. As shown in Fig. 2(a), thermalizing boundary conditions with the same temperature T0 (300K) are set at the left and right boundaries, while the top and bottom boundaries are set as specular reflecting boundary conditions (specularity = 1). A uniform heat generation over the entire simulation cell is set, heat flux $q = 1.0\times10^{15} W/m^3$.

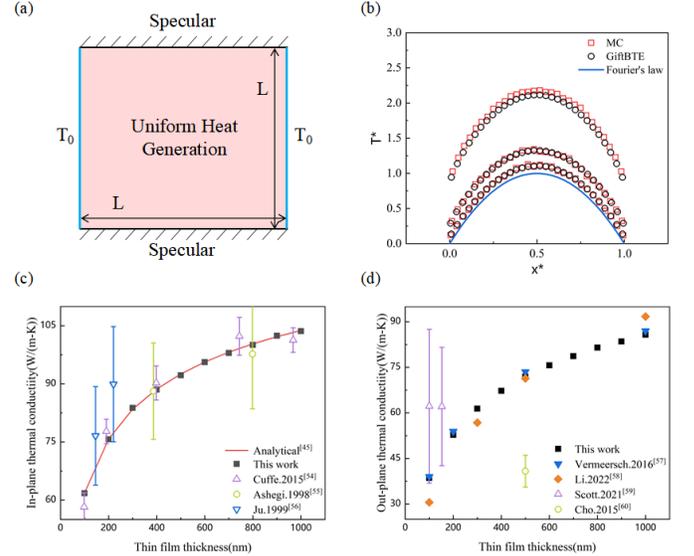

Fig. 2. (a) Schematic diagram of internal heat generation problem. (The square body with side length L is uniformly heated inside. Isothermal boundaries T$_0$=300K are added to the left and right sides, and the upper and lower boundaries are adiabatic boundaries of mirror reflection.) (b) The numerical results of internal heat generation problem verified with GiftBTE (The dimensionless temperature T* is defined as (T-T$_0$) 8κ/q/L$^2$; the dimensionless coordinate x* is defined as x/L. Among them, T$_0$ = 300 K is the cold source temperature, κ is the material bulk thermal conductivity, q is the input heat flux, and L is the structural unit characteristic size.). (c) The in-plane thermal conductivity of silicon thin films versus thickness. The black solid squares represent the results of this work, the red solid lines represent the theoretical solution results [45], the hollow graphs represent the comparison of experimental results [54], [55], [56]. (d) The out-plane thermal conductivity of silicon thin films varies with size. The black solid squares represent the results of this work, the solid graphs represent the comparison of simulation results [57], [58], and the hollow graphs represent the comparison of experimental results [59], [60].

In this study, since the chips in HBM are all Si substrates, Si is used as a material. The sets of interatomic force constants (IFCs) for Si are obtained by using VASP [61] with projector-augmented-wave (PAW) [62] pseudopotentials and Perdew–Burke–Ernzerhof (PBE) [63] exchange and correlation functionals. A 10×10×10 k-point mesh is employed for Brillouin zone sampling. Supercells of 5×5×5 and 4×4×4 are used for the 2nd-order and 3rd-order IFCs, respectively. The phonon properties of Si are calculated using the ShengBTE code [64], based on these IFCs and 30×30×30 q-point mesh. The thermal conductivity of bulk Si is calculated as 142.3 W/m-K, which is consistent with literature values [54], [60], [65], [66], [67]. The phonon grey model is used for verification and results (excluding those marked as non-grey). The grey model assumes that all phonons have the same properties. This leads to the grey model being unable to take into account the influence of phonons of different wavelengths and frequencies on heat conduction. So, the grey model is limited in that it cannot describe the anisotropy of materials and its simulation of thermal conductivity is not completely precise. However, due to its relative accuracy and low computational cost, the grey model is a widely used simplified model in previous work [38], [68], [69].

The results obtained through MC simulation are shown in Fig. 2(b). In Fig. 2(b), x* represents the dimensionless coordinate, which is defined as x/L. Fig. 2(b) shows the distribution of dimensionless lattice temperature T* in the internal heat generation problem, which is defined as $(T-T_0) \times 8k/q/L^2$. As shown in Fig. 2(b), the results for different sizes obtained by MC are consistent with the iterative solution of GiftBTE [30]. For the size of 3000 nm, we also calculated the analytical solution based on Fourier's law. According to the heat conduction equation $\frac{d^2 T}{dx^2} + \frac{q}{k} = 0$, the analytical expression of the temperature distribution can be derived as $T(x) = \frac{qL^2}{2k}(1 - \frac{x^2}{L^2}) + T_0$, which is represented by the blue solid line in Fig. 2(b). All the above results indicate the correctness of the internal heat source in the MC method.

To verify the accuracy of the model, the in-plane and out-of-plane thermal conductivities of the silicon film were calculated and compared with the previous work, as shown in Figure 1(c) and (d). The black solid squares represent the calculation results of this work, which are in good agreement with the simulation results (solid graphs) and the experimental results (hollow graphs). HBM can be regarded as a multi-layer film stack structure. The above results can effectively support the reliability of the model.

### III. RESULTS AND DISCUSSION

Fig. 3(a) shows the simplified HBM periodic structure [9], where the grey part is the Si-based chip, the blue region denotes the joint structure between chip layers (typically a lead-tin alloy), and the remaining area corresponds to the underfill layer, usually composed of an organic material. The smallest repeating structural unit in HBM is selected as our simulation domain. Due to the complex microstructure and heterogeneous composition of solder materials, along with the relatively small thickness of the joint structure, simplifications are often necessary in thermal modeling. In Fourier-based simulations, solder layers are typically treated using an equivalent thermal conductivity approach [15], [70]. For non-Fourier heat conduction scenarios, such alloy interlayers can be effectively approximated as thermal interfaces [71], [72], [73].

In our simulation, inspired by the treatment of HBM in Fourier-based models [70], the I-shaped structure is maintained, while the interlayer thermal resistance is modeled using a transmissible interface to modeling the effect of the joint structure. By referring to the joint thermal resistance in the comsol macroscopic model, the transmittance at the interface is set to 0.75 and the specularity is set to 0. Meanwhile, the temperature distribution in the case without considering the joint structure is also calculated for comparison. The bottom is set as a heat source with constant heat flux to simulate the heat generation of the chip. Since the actual heat source of the chip has a certain thickness, a 5 nm thick heat source was set in the simulation. Comprehensively consider the heat generation of the bottom logic chip and the thermal coupling of components such as the GPU, heat flux $q = 5.34 \times 10^{17} W/m^3$ [74]. The top surface is set as an isothermal boundary at 300 K to represent natural heat dissipation at room temperature. In the x and z directions, periodic boundaries are adopted, and one period is taken in the y direction. The remaining boundaries are assumed to be adiabatic. Further details can be found in Fig. 3(c) and (d). The thickness of the connector is usually half that of the chip [19], [20]. Therefore, when L = 100 nm, the joint size is a = b = 20 nm, and the joint size varies proportionally with L.

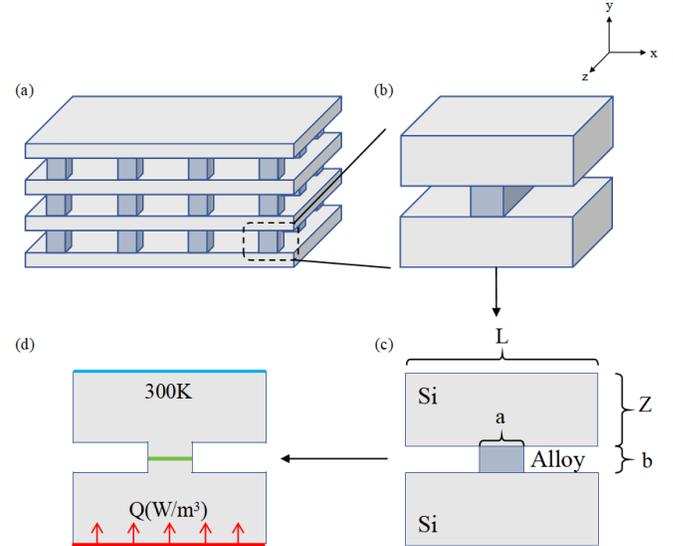

Fig. 3. (a) Schematic diagram of HBM structure, the grey film is silicon and the blue bump is alloy. (b) Structural unit of HBM. (c) Structural dimensions, including feature size L, single-layer chip thickness Z, joint width a and height b. (d) Schematic diagram of boundary conditions. The red arrow at the bottom represents the heat source of size Q, the blue line at the top represents the isothermal boundary (300K), and the green bar in the middle represents the interface.

In order to quantitatively analyze the impact of non-Fourier effects on in-chip temperature rise prediction, the temperature distribution of a single HBM structural unit with feature size L is shown in Fig. 4(a). Where the dimensionless temperature T* is defined as $(T-T_0) \times k/q/L$, and the dimensionless length Y* is defined as y/L. For Fourier solutions, there is no correlation between the temperature distribution of the structural unit and the structural feature size. However, when the non-fourier heat transport is considered, the simulation results from the MC method differ significantly from those predicted by the Fourier model as the structural feature size decreases. Notably, when the characteristic size of the HBM structural unit is reduced to 100 nm, the highest absolute temperature predicted by the two methods is 95.2 °C for the MC simulation and 35.4 °C for the Fourier solution, as shown in Fig. 4(b). Due to the interfacial thermal resistance, both of them exhibit a significant temperature jump near Y*=0.5. Since the characteristic size of the system has reached a magnitude comparable to the average free path of phonons, phonon ballistic transport and phonon boundary scattering deteriorate the thermal conductivity of the system [75], [76]. This leads to the junction temperature obtained by MC being much higher than the Fourier solution which is dominated by diffusion phonon transport. The result of MC is 59.8 ℃ higher than that predicted by Fourier solution, as shown by the red solid line and the black solid line in the figure. Even when the film thermal conductivity obtained by considering phonon scattering at boundaries is substituted into the Fourier model, as indicated by the dashed black line, there is still a certain gap compared with the results of MC, which cannot be ignored in chip design. This discrepancy arises because Fourier's law or even its modified versions, which assumes diffusive heat transport, fails to accurately describe phonon behavior at the nanoscale. The Monte Carlo (MC) method was adopted as a numerical solver for the BTE, providing a more physically accurate prediction of nanoscale heat transport [77], [78].

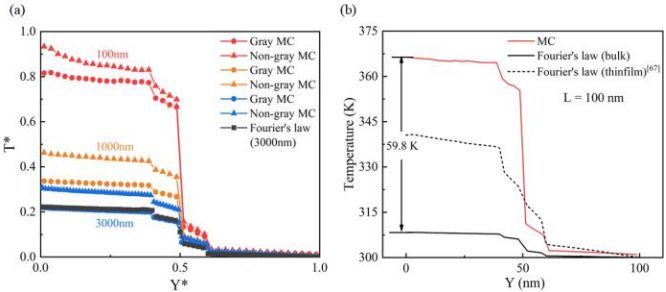

Fig. 4. (a) One-dimensional temperature distribution in y direction in a single structural unit, the Fourier method uses dashed lines for description, while the MC method uses solid lines. The gray model in the MC uses circular symbols, and the non-gray model uses triangular symbols. Different colors represent different sizes of L, and the values of L are marked in the figure. (The dimensionless temperature T* is defined as (T-T0) κ/q/L, the dimensionless coordinate Y* is defined as y/L. Among them, T0 = 300 K is the cold source temperature, κ is the material bulk thermal conductivity, q is the input heat flux, and L is the structural unit characteristic size). (b) Comparison of temperature distribution in y direction within a single structural unit at L = 100 nm. The solid red line represents the results of the MC method, the solid black line is the Fourier result under the bulk thermal conductivity model, and the dashed black line is the Fourier result under the thin film thermal conductivity model [67].

In Figure 5, two simplified MC models are considered. To observe the temperature of the joint part in the structure, the joint structure in the region from Y = 40 nm to Y = 60 nm is retained in Fig 5(a). Fig 5(b) simplifies the entire joint structure in the region from Y = 40 nm to Y = 60 nm into an interface. There is no obvious temperature difference between the two simplified models. In contrast to MC, Fig 5(c) is the result of the Fourier calculation. It can be observed that in the MC simulation, the junction temperature within the chip is higher, and the temperature distribution is more uneven with hot spots [79], [80]. This is because, unlike the isotropic thermal conductivity assumed in the Fourier method, the Monte Carlo (MC) method explicitly accounts for phonon scattering at boundaries. These scattering events cause an uneven phonon distribution, leading to regions with higher phonon density and consequently localized temperature increases.

It is worth noting in Fig 4(a) that when the silicon phonon non-grey model is used as input, it introduces a more significant non-Fourier effect. The grey model assumes that the free-path of all phonons are equal to an average value, this will lead to ignoring the effect of long free-path phonons. When a more detailed non-grey model is adopted, these phonons with long free-path generate more scattering events at the boundaries, thereby generating additional non-Fourier effects that cannot be observed in the grey model. This is reflected in two key aspects: first, the temperature difference between the non-Fourier model and the Fourier method is further amplified; second, non-Fourier effects influence a broader range of sizes. This implies that non-Fourier heat transport remains significant even at larger sizes, and Fourier's law fails to accurately predict the chip temperature distribution. In the HBM system, multi-layer chip stacking introduces additional boundaries, leading to increased phonon scattering, which exacerbates heat dissipation within the chip. Therefore, it is essential to consider non-Fourier heat transport in future chip designs.

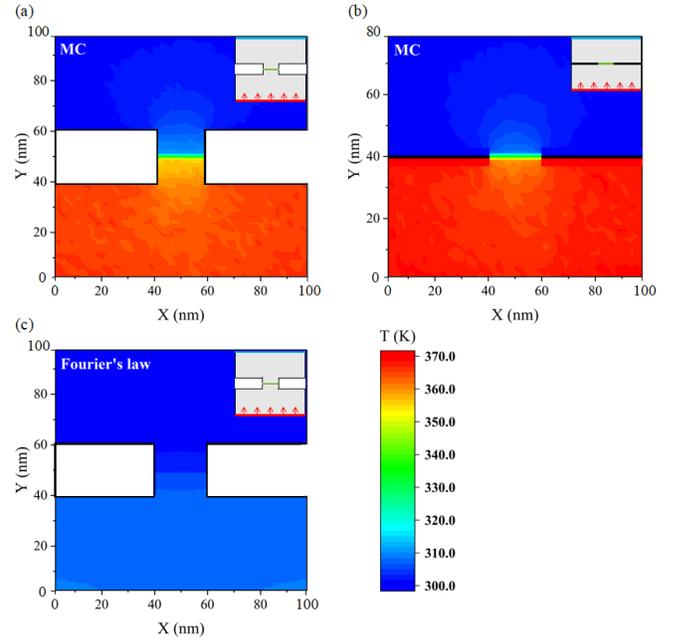

Fig. 5. Temperature distribution cloud map. (The illustration shows the schematic diagram of the corresponding structural unit: the bottom is the heat flow source, the middle is the interface structure, and the top is the isothermal boundary.) (a) Phonon MC solution, L=100 nm (I-shape); (b) Phonon MC solution, L=100 nm (Square-shape); (c) Fourier's law, L=100 nm.

It can be clearly observed in Fig. 4(a) that there is a noticeable temperature jump in the range of $0.4 \leq Y^* \leq 0.6$. This is due to the fact that the interlayer thermal resistance between the chip layers limits the heat transport. In the thermal design of 3D chips, the interface thermal resistance can be improved by changing materials, improving bonding processes, etc [81], [82], [83]. These methods essentially increase the interface transmittance [84]. Under different transmissible interface conditions, the average temperature at the interface were calculated. The influence of interlayer thermal resistance on heat transport primarily depends on two parameters: specularity and transmittance. Previous studies have shown that specularity has a relatively minor impact on thermal transport. At room temperature, assuming fully diffuse scattering (specularity = 0) yields results that are sufficiently accurate for practical purposes [85], [86].

In Fig. 6(a), we calculated the changes of temperature with the transmittance when the specularity is set as 0. It is evident that the transmittance of the transmissible interface plays a decisive role in the heat transport at the joint. As the transmittance decreases, the temperature of the chip increases sharply. With a transmittance of 0.2, the temperature rise is 56.6 °C higher than when the transmittance is 0.8, and Fig. 6(b) shows that the interlayer thermal resistance varies by 8 times under the two transmittance, highlighting the crucial role of interlayer thermal resistance in the heat dissipation of HBM. The interlayer thermal resistance is obtained by dividing the temperature difference of the 20 nm joint by the average heat flow. To verify the reliability of this thermal resistance, the thermal resistances of the Pb-Sn and Si-Sn alloys were plotted in Fig. 6(b). The Pb-Sn alloy is a commonly used solder for HBM [87], and the thickness of the Si-Sn alloy film is consistent with this simulation, which is 20 nm [88]. In Fig. 6(d), it can be visually observed that there is a uniform temperature gradient at the joint under high transmittance. While in Fig. 6(c), the low transmittance hinders heat transfer and causes a temperature jump at the joint, which is harmful to the thermal management of the chips. Therefore, high transmittance joint materials and bonding processes are urgently needed in HBM thermal management.

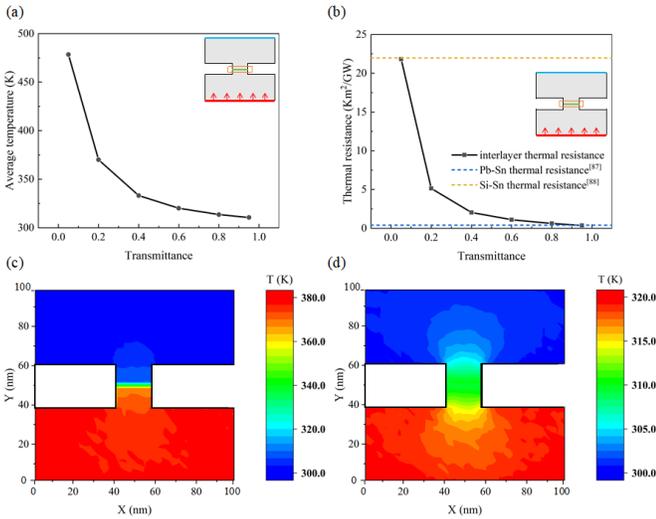

Fig. 6. (a) Variation of temperature at the interface with the interfacial transmittance (the interfacial specularity is 0, the sampling area is shown in the illustration). (b) Variation of interlayer thermal resistance with the interfacial transmittance (the interfacial specularity is 0, the area for calculating thermal resistance is shown in the illustration). For reference and comparison, the thermal resistance of some common solder alloys is shown in the figure with dashed lines. The thermal resistance of 35Pb-65Sn calculated using the bulk thermal conductivity [87] and the thermal resistance of 41Si-59Sn calculated using the 20nm film thermal conductivity [88]. (c) The temperature distribution of structural units with interfacial transmittance of 0.2. (d) The temperature distribution of structural units with interfacial transmittance of 0.8.

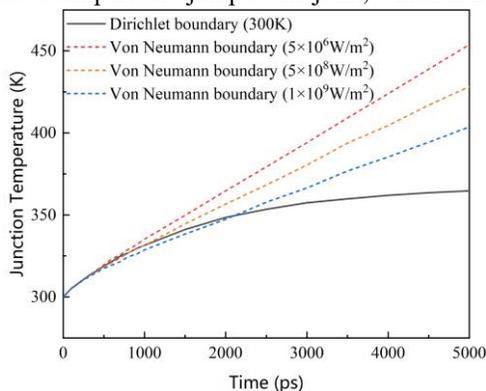

Fig. 7. Comparison of junction temperatures in HBM under different boundary conditions. The solid line represents the Dirichlet boundary condition (T = 300K), the dashed line represents the Von Neumann boundary condition ($Q_{cooling}$ = 5×10$^6$/5×10$^8$/1×10$^9$ (W/m$^2$))

Due to the actual cooling of the chip system is not ideal, the junction temperatures of the HBM system under different boundary conditions were calculated, as shown in Fig 7. When setting the Dirichlet boundary conditions, the system is ideally cooled, and the junction temperature tends to stabilize over time. When setting the Von Neumann boundary conditions, usually, the heat generation power of the chip is greater than the heat dissipation power, energy is continuously input into the system. As a result, the junction temperature increases monotonically over time. As the cooling power increases, the junction temperature within the HBM system is reduced.

## IV. Conclusion

The phonon MC is used to simulate the temperature distribution in the 3D structure of HBM memory chip. The numerical results show that the traditional Fourier model significantly underestimates the junction temperature of the chip at small scales, reaching 59.8 °C. This difference will intensify as the size decreases. Using the grey model, obvious non-Fourier effects can be observed below 3 $\mu m$. Using the non-grey model, due to the long free-path phonon effect, this critical size is expected to expand to more than 10 $\mu m$, which is the size achievable in chip design.

Additionally, the chip temperature is compared under different interface conditions. The results show that the phonon transmittance at the joint is a critical factor influencing the heat dissipation of the chip. When the interface transmittance decreases from 0.8 to 0.2, the interlayer thermal resistance will increase by 8 times, which can generate a temperature rise of 56.6 °C. Meanwhile, the low transmittance will also cause a large temperature jump at the interface, which will lead to a large temperature difference within the joint, deteriorate its performance and be unfavorable for the thermal management of HBM. Materials and bonding processes with high phonon transmittance can be effectively applied in industry, significantly improving the heat dissipation of HBM. This study provides more reliable guidance for the thermodynamic analysis of HBM systems.

## V. Reference

[1] C. Y. Lee, C. H. Won, S. Jung, E. S. Jung, T. M. Choi, H. R. Lee, et al. 3D Integrated Process and Hybrid Bonding of High Bandwidth Memory (HBM). Electronic Materials Letters, 2025: 1-25

[2] K. Kara, C. Hagleitner, D. Diamantopoulos, D. Syrivelis, G. Alonso. High Bandwidth Memory on FPGAs: A Data Analytics Perspective, in: 2020 30th International Conference on Field-Programmable Logic and Applications (FPL), Gothenburg, Sweden, IEEE, 2020: 1-8

[3] J. H. Chae. High-Bandwidth and Energy-Efficient Memory Interfaces for the Data-Centric Era: Recent Advances, Design Challenges, and Future Prospects. IEEE Open Journal of the Solid-State Circuits Society, 2024, 4: 252-264

[4] H. Jun, S. Nam, H. Jin, J. C. Lee, Y. J. Park, J. J. Lee. High-Bandwidth Memory (HBM) Test Challenges and Solutions. IEEE Design & Test, 2017, 34(1): 16-25

[5] Joonyoung Kim, Younsu Kim. HBM: Memory solution for bandwidth-hungry processors, in: 2014 IEEE Hot Chips 26 Symposium (HCS), Cupertino, CA, USA, IEEE, 2014: 1-24

[6] A. K. Kabat, S. Pandey, V. T. Gopalakrishnan. Performance evaluation of High Bandwidth Memory for HPC Workloads, in: 2022 IEEE 35th

International System-on-Chip Conference (SOCC), Belfast, United Kingdom, IEEE, 2022: 1-6
[7] S. Ha, S. Lee, Gh. Bae, Ds. Lee, S. H. Kim, Bw. Woo, et al. Reliability Characterization of HBM featuring HK+MG Logic Chip with Multi-stacked DRAMs, in: 2023 IEEE International Reliability Physics Symposium (IRPS), Monterey, CA, USA, IEEE, 2023: 1-7
[8] K. Cho, H. Lee, H. Kim, S. Choi, Y. Kim, J. Lim, et al. Design optimization of high bandwidth memory (HBM) interposer considering signal integrity, in: 2015 IEEE Electrical Design of Advanced Packaging and Systems Symposium (EDAPS), Seoul, South Korea, IEEE, 2015: 15-18
[9] H. Jun, J. Cho, K. Lee, H. Y. Son, K. Kim, H. Jin, et al. HBM (High Bandwidth Memory) DRAM Technology and Architecture, in: 2017 IEEE International Memory Workshop (IMW), Monterey, CA, USA, IEEE, 2017: 1-4
[10] M. Zhu, Y. Zhuo, C. Wang, W. Chen, Y. Xie. Performance Evaluation and Optimization of HBM-Enabled GPU for Data-Intensive Applications. IEEE Transactions on Very Large Scale Integration (VLSI) Systems, 2018, 26(5): 831-840
[11] K. Kim, M. jae Park. Present and Future, Challenges of High Bandwidth Memory (HBM), in: 2024 IEEE International Memory Workshop (IMW), Seoul, Korea, Republic of, IEEE, 2024: 1-4
[12] C. Zhang, Y. Liu, Q. Chen. Neural Network Surrogate Model for Junction Temperature and Hotspot Position in 3D Multi-Layer High Bandwidth Memory (HBM) Chiplets under Varying Thermal Conditions. arXiv, 2025
[13] S. K. Rajan, A. Kaul, T. Sarvey, G. S. May, M. S. Bakir. Design Considerations, Demonstration, and Benchmarking of Silicon Microcold Plate and Monolithic Microfluidic Cooling for 2.5D ICs, in: 2021 IEEE 71st Electronic Components and Technology Conference (ECTC), San Diego, CA, USA, IEEE, 2021: 1418-1426
[14] K. Son, J. Park, S. Kim, B. Sim, K. Kim, S. Choi, et al. Thermal Analysis of High Bandwidth Memory (HBM)-GPU Module considering Power Consumption, in: 2023 IEEE Electrical Design of Advanced Packaging and Systems (EDAPS), Rose-Hill, Mauritius, IEEE, 2023: 1-3
[15] T. Kim, J. Lee, Y. Kim, H. Park, H. Hwang, J. Kim, et al. Thermal Improvement of HBM with Joint Thermal Resistance Reduction for Scaling 12 Stacks and Beyond, in: 2023 IEEE 73rd Electronic Components and Technology Conference (ECTC), Orlando, FL, USA, IEEE, 2023: 767-771
[16] Y. Zhang, T. E. Sarvey, M. S. Bakir. Thermal Evaluation of 2.5-D Integration Using Bridge-Chip Technology: Challenges and Opportunities. IEEE Transactions on Components, Packaging and Manufacturing Technology, 2017, 7(7): 1101-1110
[17] S. K. Rajan, A. Kaul, T. Sarvey, G. S. May, M. S. Bakir. Design Considerations, Demonstration, and Benchmarking of Silicon Microcold Plate and Monolithic Microfluidic Cooling for 2.5D ICs, in: 2021 IEEE 71st Electronic Components and Technology Conference (ECTC), San Diego, CA, USA, IEEE, 2021: 1418-1426
[18] K. Son, S. Kim, H. Park, S. Kim, K. Kim, S. Park, et al. A Novel Through Mold Plate (TMP) for Signal and Thermal Integrity Improvement of High Bandwidth Memory (HBM), in: 2020 IEEE MTT-S International Conference on Numerical Electromagnetic and Multiphysics Modeling and Optimization (NEMO), Hangzhou, China, IEEE, 2020: 1-4
[19] K. Son, S. Kim, S. Park, H. Park, K. Kim, T. Shin, et al. Design and Analysis of Thermal Transmission Line based Embedded Cooling Structures for High Bandwidth Memory Module and 2.5D/3D ICs, in: 2020 IEEE Electrical Design of Advanced Packaging and Systems (EDAPS), Shenzhen, China, IEEE, 2020: 1-3
[20] K. Son, S. Kim, H. Park, T. Shin, K. Kim, M. Kim, et al. Thermal and Signal Integrity Co-Design and Verification of Embedded Cooling Structure With Thermal Transmission Line for High Bandwidth Memory Module. IEEE Transactions on Components, Packaging and Manufacturing Technology, 2022, 12(9): 1542-1556
[21] T. W. Wei, H. Oprins, V. Cherman, I. De Wolf, G. Van Der Plas, E. Beyne, et al. Thermal Analysis of Polymer 3D Printed Jet Impingement Coolers for High Performance 2.5D Si Interposer Packages, 2019 18th IEEE Intersociety Conference on Thermal and Thermomechanical Phenomena in Electronic Systems (ITherm), Las Vegas, NV, USA, IEEE, 2019: 1243-1252
[22] Pan D. K., Zong Z. C., Yang N., School of Energy and Power Engineering, Huazhong University of Science and Technology, Wuhan 430074, China, College of New Energy, China University of Petroleum (East China), Qingdao 266580, China. Phonon weak couplings in nanoscale thermophysics. Acta Physica Sinica, 2022, 71(8): 086302
[23] C. Deng, Y. Huang, M. An, N. Yang. Phonon weak couplings model and its applications: A revisit to two-temperature non-equilibrium transport. Materials Today Physics, 2021, 16: 100305
[24] N. Yang, G. Zhang, B. Li. Violation of Fourier's law and anomalous heat diffusion in silicon nanowires. Nano Today, 2010, 5(2): 85-90
[25] Xiao Y, Chen Q, Ma D, et al. Phonon Transport within Periodic Porous Structures — From Classical Phonon Size Effects to Wave Effects. ES Materials & Manufacturing, 2019, 5, 2–18
[26] G. Chen. Non-Fourier phonon heat conduction at the microscale and nanoscale. Nature Reviews Physics, 2021, 3(8): 555-569
[27] M. Wang, N. Yang, Z. Y. Guo. Non-Fourier heat conductions in nanomaterials. Journal of Applied Physics, 2011, 110(6): 064310
[28] C. Landon, L. Jiang, D. Pantuso, I. Meric, K. Komeyli, J. Hicks, et al. Localized thermal effects in Gate-all-around devices, in: 2023 IEEE International Reliability Physics Symposium (IRPS), Monterey, CA, USA, IEEE, 2023: 1-5
[29] X. Chang, H. Oprins, M. Lofrano, V. Cherman, B. Vermeersch, J. D. Fortuny, et al. Calibrated fast thermal calculation and experimental characterization of advanced BEOL stacks, in: 2023 IEEE International Interconnect Technology Conference (IITC) and IEEE Materials for Advanced Metallization Conference (MAM)(IITC/MAM), Dresden, Germany, IEEE, 2023: 1-3
[30] Y. Hu, R. Jia, J. Xu, Y. Sheng, M. Wen, J. Lin, et al. GiftBTE: an efficient deterministic solver for non-gray phonon Boltzmann transport equation. Journal of Physics: Condensed Matter, 2024, 36(2): 025901
[31] Y. Xia, X. Zhang, A. Wang, Y. Sheng, H. Xie, H. Bao. Critical factors influencing electron and phonon thermal conductivity in metallic materials using first-principles calculations. Journal of Physics: Condensed Matter, 2025, 37(5): 055701
[32] Y. Sheng, S. Wang, Y. Hu, J. Xu, Z. Ji, H. Bao. Integrating First-Principles-Based Non-Fourier Thermal Analysis Into Nanoscale Device Simulation. IEEE Transactions on Electron Devices, 2024, 71(3): 1769-1775
[33] J. Xu, Y. Hu, H. Bao. Quantitative Analysis of Nonequilibrium Phonon Transport Near a Nanoscale Hotspot. Physical Review Applied, 2023, 19(1): 014007
[34] Bao H, Chen J, Gu X, et al. A review of simulation methods in micro/nanoscale heat conduction. ES Energy & Environment, 2018, 1(84): 16-55.
[35] D. S. Tang, B. Y. Cao. Phonon thermal transport and its tunability in GaN for near-junction thermal management of electronics: A review. International Journal of Heat and Mass Transfer, 2023, 200: 123497
[36] H. L. Li, Y. Shen, Y. C. Hua, S. L. Sobolev, B. Y. Cao. Hybrid Monte Carlo-Diffusion Studies of Modeling Self-Heating in Ballistic-Diffusive Regime for Gallium Nitride HEMTs. Journal of Electronic Packaging, 2023, 145(1): 011203
[37] Z. L. Tang, Y. Shen, H. L. Li, B. Y. Cao. Topology optimization for near-junction thermal spreading of electronics in ballistic-diffusive regime. iScience, 2023, 26(7): 107179
[38] Y. Shen, H. A. Yang, B. Y. Cao. Near-junction phonon thermal spreading in GaN HEMTs: A comparative study of simulation techniques by full-band phonon Monte Carlo method. International Journal of Heat and Mass Transfer, 2023, 211: 124284
[39] Q. Hao, H. Zhao, Y. Xiao, Q. Wang, X. Wang. Hybrid Electrothermal Simulation of a 3-D Fin-Shaped Field-Effect Transistor Based on GaN Nanowires. IEEE Transactions on Electron Devices, 2018, 65(3): 921-927
[40] S. V. J. Narumanchi, J. Y. Murthy, C. H. Amon. Boltzmann transport equation-based thermal modeling approaches for hotspots in microelectronics. Heat and Mass Transfer, 2006, 42(6): 478-491
[41] J. P. M. Péraud, N. G. Hadjiconstantinou. Efficient simulation of multidimensional phonon transport using energy-based variance-reduced Monte Carlo formulations. Physical Review B, 2011, 84(20): 205331
[42] P. Allu, S. Mazumder. Hybrid ballistic–diffusive solution to the frequency-dependent phonon Boltzmann Transport Equation. International Journal of Heat and Mass Transfer, 2016, 100: 165-177
[43] S. Mazumder, A. Majumdar. Monte Carlo Study of Phonon Transport in Solid Thin Films Including Dispersion and Polarization. Journal of Heat Transfer, 2001, 123(4): 749-759
[44] G. Xie, D. Ding, G. Zhang. Phonon coherence and its effect on thermal conductivity of nanostructures. Advances in Physics: X, 2018, 3(1): 1480417

[45] G. Chen. Nanoscale energy transport and conversion: a parallel treatment of electrons, molecules, phonons, and photons. Oxford: Oxford University Press, 2005

[46] D. Lacroix, K. Joulain, D. Lemonnier. Monte Carlo transient phonon transport in silicon and germanium at nanoscales. Physical Review B, 2005, 72(6): 064305

[47] N. G. Hadjiconstantinou, A. L. Garcia, M. Z. Bazant, G. He. Statistical error in particle simulations of hydrodynamic phenomena. Journal of Computational Physics, 2003, 187(1): 274-297

[48] J. P. M. Péraud, N. G. Hadjiconstantinou. An alternative approach to efficient simulation of micro/nanoscale phonon transport. Applied Physics Letters, 2012, 101(15): 153114

[49] Q. Hao, G. Chen, M. S. Jeng. Frequency-dependent Monte Carlo simulations of phonon transport in two-dimensional porous silicon with aligned pores. Journal of Applied Physics, 2009, 106(11): 114321

[50] S. Wolf, N. Neophytou, H. Kosina. Thermal conductivity of silicon nanomeshes: Effects of porosity and roughness. Journal of Applied Physics, 2014, 115(20): 204306

[51] D. S. Tang, Y. C. Hua, B. Y. Cao. Thermal wave propagation through nanofilms in ballistic-diffusive regime by Monte Carlo simulations. International Journal of Thermal Sciences, 2016, 109: 81-89

[52] D. S. Tang, B. Y. Cao. Superballistic characteristics in transient phonon ballistic-diffusive transport. Applied Physics Letters, 2017, 111(11): 113109

[53] A. Pathak, A. Pawnday, A. P. Roy, A. J. Aref, G. F. Dargush, D. Bansal. MCBTE: A variance-reduced Monte Carlo solution of the linearized Boltzmann transport equation for phonons. Computer Physics Communications, 2021, 265: 108003

[54] J. Cuffe, J. K. Eliason, A. A. Maznev, K. C. Collins, J. A. Johnson, A. Shchepetov, et al. Reconstructing phonon mean-free-path contributions to thermal conductivity using nanoscale membranes. Physical Review B, 2015, 91(24): 245423

[55] M. Asheghi, Y. K. Leung, S. S. Wong, K. E. Goodson. Phonon-boundary scattering in thin silicon layers. Applied Physics Letters, 1997, 71(13): 1798-1800

[56] Y. S. Ju, K. E. Goodson. Phonon scattering in silicon films with thickness of order 100 nm. Applied Physics Letters, 1999, 74(20): 3005-3007

[57] B. Vermeersch, J. Carrete, N. Mingo. Cross-plane heat conduction in thin films with ab-initio phonon dispersions and scattering rates. Applied Physics Letters, 2016, 108(19): 193104

[58] R. Li, J. X. Wang, E. Lee, T. Luo. Physics-informed deep learning for solving phonon Boltzmann transport equation with large temperature non-equilibrium. npj Computational Materials, 2022, 8(1): 29

[59] E. A. Scott, C. Perez, C. Saltonstall, D. P. Adams, V. Carter Hodges, M. Asheghi, et al. Simultaneous thickness and thermal conductivity measurements of thinned silicon from 100 nm to 17 μm. Applied Physics Letters, 2021, 118(20): 202108

[60] J. Cho, D. Francis, P. C. Chao, M. Asheghi, K. E. Goodson. Cross-Plane Phonon Conduction in Polycrystalline Silicon Films. Journal of Heat Transfer, 2015, 137(7): 071303

[61] G. Kresse, J. Furthmüller. Efficient iterative schemes for ab initio total-energy calculations using a plane-wave basis set. Physical Review B, 1996, 54(16): 11169-11186

[62] P. E. Blöchl. Projector augmented-wave method. Physical Review B, 1994, 50(24): 17953-17979

[63] J. P. Perdew, K. Burke, M. Ernzerhof. Generalized Gradient Approximation Made Simple. Physical Review Letters, 1996, 77(18): 3865-3868

[64] W. Li, J. Carrete, N. A. Katcho, N. Mingo. ShengBTE: A solver of the Boltzmann transport equation for phonons. Computer Physics Communications, 2014, 185(6): 1747-1758

[65] E. Chávez-Ángel, J. S. Reparaz, J. Gomis-Bresco, M. R. Wagner, J. Cuffe, B. Graczykowski, et al. Reduction of the thermal conductivity in free-standing silicon nano-membranes investigated by non-invasive Raman thermometry. APL Materials, 2014, 2(1): 012113

[66] X. P. Luo, H. L. Yi. A discrete unified gas kinetic scheme for phonon Boltzmann transport equation accounting for phonon dispersion and polarization. International Journal of Heat and Mass Transfer, 2017, 114: 970-980

[67] X. Ran, Y. Huang, M. Wang. A hybrid Monte Carlo-discrete ordinates method for phonon transport in micro/nanosystems with rough interfaces. International Journal of Heat and Mass Transfer, 2023, 201: 123624

[68] K. Etessam-Yazdani, Yizhang Yang, M. Asheghi. Ballistic phonon transport and self-heating effects in strained-silicon transistors. IEEE Transactions on Components and Packaging Technologies, 2006, 29(2): 254-260

[69] H. L. Li, J. Shiomi, B. Y. Cao. Ballistic-Diffusive Heat Conduction in Thin Films by Phonon Monte Carlo Method: Gray Medium Approximation Versus Phonon Dispersion. Journal of Heat Transfer, 2020, 142(11): 112502

[70] T. Kim, J. Lee, J. Kim, E. C. Lee, H. Hwang, Y. Kim, et al. Thermal Modeling and Analysis of High Bandwidth Memory in 2.5D Si-interposer Systems, in: 2022 21st IEEE Intersociety Conference on Thermal and Thermomechanical Phenomena in Electronic Systems (iTherm), San Diego, CA, USA, IEEE, 2022: 1-5

[71] Q. Li, F. Liu, Y. Liu, T. Wang, X. Wang, B. Sun. Effect of the alloyed interlayer on the thermal conductance of Al/GaN interface. Journal of Applied Physics, 2023, 134(23): 230901

[72] Q. Li, F. Liu, S. Hu, H. Song, S. Yang, H. Jiang, et al. Inelastic phonon transport across atomically sharp metal/semiconductor interfaces. Nature Communications, 2022, 13(1): 4901

[73] P. E. Hopkins, P. M. Norris, R. J. Stevens, T. E. Beechem, S. Graham. Influence of Interfacial Mixing on Thermal Boundary Conductance Across a Chromium/Silicon Interface. Journal of Heat Transfer, 2008, 130(6): 062402

[74] A. Patel, K. Yogi, G. Sahu, T. Wei. Multi-chip Jet impingement cooling for heat dissipation in 2.5D integrated system with 1 kW+ thermal design power. International Journal of Heat and Mass Transfer, 2025, 244: 126978

[75] H. Meng, D. Ma, X. Yu, L. Zhang, Z. Sun, N. Yang. Thermal conductivity of molybdenum disulfide nanotube from molecular dynamics simulations. International Journal of Heat and Mass Transfer, 2019, 145: 118719

[76] H. Meng, S. Maruyama, R. Xiang, N. Yang. Thermal conductivity of one-dimensional carbon-boron nitride van der Waals heterostructure: A molecular dynamics study. International Journal of Heat and Mass Transfer, 2021, 180: 121773

[77] L. Yang, A. J. Minnich. Thermal transport in nanocrystalline Si and SiGe by ab initio based Monte Carlo simulation. Scientific Reports, 2017, 7(1): 44254

[78] L. Yang, Y. Jiang, Y. Zhou. Quantitatively predicting modal thermal conductivity of nanocrystalline Si by full-band Monte Carlo simulations. Physical Review B, 2021, 104(19): 195303

[79] C. Zhang, D. Ma, M. Shang, X. Wan, J. T. Lü, Z. Guo, et al. Graded thermal conductivity in 2D and 3D homogeneous hotspot systems. Materials Today Physics, 2022, 22: 100605

[80] Wu Z. P., Zhang C., Hu S. Q., Ma D. K., Yang N., School of Energy and Power Engineering, Huazhong University of Science and Technology, Wuhan 430074, China, et al. Graded thermal conductivity in nano "hot spot" systems. Acta Physica Sinica, 2023, 72(18): 184401

[81] W. Xing, Y. Xu, C. Song, T. Deng. Recent Advances in Thermal Interface Materials for Thermal Management of High-Power Electronics. 2022

[82] Zong Z. C., Pan D. K., Deng S. C., Wan X., Yang L. N., Ma D. K., et al. Mixed mismatch model predicted interfacial thermal conductance of metal/semiconductor interface. Acta Physica Sinica, 2023, 72(3): 034401

[83] Z. Zong, S. Deng, Y. Qin, X. Wan, J. Zhan, D. Ma, et al. Enhancing the interfacial thermal conductance of Si/PVDF by strengthening atomic couplings. Nanoscale, 2023, 15(40): 16472-16479

[84] L. Chen, Z. Huang, S. Kumar. Phonon transmission and thermal conductance across graphene/Cu interface. Appl. Phys. Lett., 2025

[85] A. M. Marconnet, M. Asheghi, K. E. Goodson. From the Casimir Limit to Phononic Crystals: 20 Years of Phonon Transport Studies Using Silicon-on-Insulator Technology. Journal of Heat Transfer, 2013, 135(6): 061601

[86] D. Gelda, M. G. Ghossoub, K. Valavala, J. Ma, M. C. Rajagopal, S. Sinha. Specularity of longitudinal acoustic phonons at rough surfaces. Physical Review B, 2018, 97(4): 045429

[87] Y. Ocak, S. Aksöz, N. Maraşlı, E. Çadırlı. Dependency of thermal and electrical conductivity on temperature and composition of Sn in Pb–Sn alloys. Fluid Phase Equilibria, 2010, 295(1): 60-67

[88] S. N. Khatami, Z. Aksamija. Lattice Thermal Conductivity of the Binary and Ternary Group-IV Alloys Si-Sn, Ge-Sn, and Si-Ge-Sn. Physical Review Applied, 2016, 6(1): 014015